\documentclass[prl,showpacs,floatfix, superscriptaddress, preprint]{revtex4-1}
\usepackage{graphicx}%
\usepackage{dcolumn}
\usepackage{amsmath}
\begin{document}

\title{Dynamic Solidification in Nanoconfined Water Films}
\author{Shah H. Khan}
\affiliation{Department of Physics and Astronomy, Wayne State University, Detroit, MI 48201, USA}
\author{George Matei}
\affiliation{Department of Physics and Astronomy, Wayne State University, Detroit, MI 48201, USA}
\author{Shivprasad Patil}
\affiliation{Indian Institute of Science Education and Research, Pune 411021, India}
\author{Peter M. Hoffmann}
\email{hoffmann@wayne.edu}
\affiliation{Department of Physics and Astronomy, Wayne State University, Detroit, MI 48201, USA}
\date{\today}

\begin{abstract}
The mechanical properties of nanoconfined water layers are still poorly understood and continue to create considerable controversy, despite their importance for biology and nanotechnology. Here, we report on dynamic nanomechanical measurements of water films compressed down to a few single molecular layers. We show that the mechanical properties of nanoconfined water layers change dramatically with their dynamic state. In particular, we observed a sharp transition from viscous to elastic response even at extremely slow compression rates, indicating that mechanical relaxation times increase dramatically once water is compressed to less than 3-4 molecular layers.
\end{abstract}
\maketitle

Water is the fundamental solvent of all living organisms \cite{fin96} and plays a crucial role in macromolecular structure formation. In nanotribology and nanofluidics \cite{hol06}, the behavior of molecularly-thin water films is also crucial. Yet, the influence of the nanomechanical dynamics of water is largely unexplored.  Moreover, nanomechanical measurements of water have produced contradictory results. Controversial questions include the properties of water close to hydrophobic surfaces \cite{poy06}, the evidence (or lack thereof) of a sharp increase in viscosity upon confinement \cite{ant01,jef04,li07,rav01,goe07}, and the question of the no-slip boundary condition \cite{hon07}.

Atomic Force Microscopy (AFM) and Surface Force Apparatus (SFA) measurements suggest that water layers confined between hydrophilic surfaces assume spontaneous order \cite{isr83,ant01,jef04,li07,uch05} and exhibit sharp increases in effective viscosity, relaxation times, and elasticity \cite{ant01,jef04,li07}. However, other measurements indicate that water under similar circumstances shows little change in effective viscosity \cite{rav01}. It is also not clear if layering influences only the elastic response of the liquid or both the viscous and elastic response \cite{jef04, pat06, maa06, osh06, kag08}. Recent measurements have shown that nanoconfined liquids can exhibit sharp changes in viscoelastic properties in response to mild changes in their dynamical state \cite{pat06,hof09,zhu03}. To resolve these issues, it is therefore imperative to carefully measure the elastic and viscous response of nanoconfined water layers under different dynamic conditions.

\begin{figure}\centerline{\includegraphics[width=83mm, clip,
keepaspectratio]{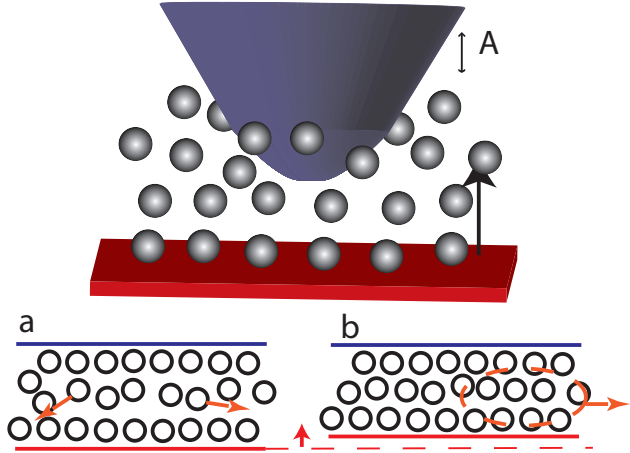}} \caption{(color online) Schematic showing the mica substrate approach the AFM tip, while water molecules are confined in the gap. The AFM tip is oscillated at an amplitude smaller than the molecular size. a. The confined water layer is between three and four molecular layers thick, and does not exhibit ordered layers. Therefore, molecules have sufficient space to easily diffuse. (tip represented by top, blue line) b. Water layer is further compressed until the gap corresponds to three molecular layers. Molecules are ordered in the vertical direction into layers, and cannot easily move individually. Squeeze-out of a layer requires collective motion of many molecules.}\label{scheme}
\end{figure}

We used a small-amplitude (A=1$\textrm{\AA}$) AFM technique\cite{pat05}, developed in our lab, to perform linear viscoelastic measurements of molecularly confined ultrapure water layers at extremely slow loading rates (Schematic see Fig. \ref{scheme}). Although we used ultrapure water, there could be a substantial amount of ions in solution originating from the freshly cleaved mica surface. Measurements were performed far below the resonance to ascertain well-behaved phase behavior of the cantilever motion. This ensures that phase changes corresponded to the dissipative behavior of the liquid and not the complicated phase behavior of the cantilever. The loading rate was controlled by the approach speed of an atomically flat mica surface towards a silicon AFM tip from 2 $\textrm{\AA}$/s to 14 $\textrm{\AA}$/s. At these speeds, the tip takes between 1.25 s to 0.18 s to traverse one molecular layer of water (width 2.5 \textrm{\AA}). This is extremely slow compared to molecular re-arrangement times. For the measurements, we immersed the cantilever and substrate in a liquid cell filled with pure water. We continuously measured the cantilever amplitude and phase using a very sensitive fiber interferometer while the sample was approached until contact with the mica surface occurred. Contact was determined from a combination of observing the amplitude approach a minimum close to zero, a strong static deflection of the cantilever, and a large change in the phase. From the phase and amplitude of the cantilever, we calculated the effective stiffness, according to $k=k_L (\frac{A_0}{A}\cos\phi-1)$, where $k_L$ is the cantilever stiffness, $A_0$ the drive piezo amplitude, $A$ the measured cantilever amplitude and $\phi$  the cantilever phase; and the damping coefficient according to $\gamma=-\frac{k_L A_0}{A \omega}\sin\phi$. The mechanical relaxation time \cite{jef04} of the confined water layer is given by  $t_R=\frac{k}{\gamma \omega^2}$. Based on the viscoelastic Maxwell model, this relaxation time corresponds to the time during which most of the stress dissipates when an external strain is imposed. In liquids, stresses dissipate quickly, whereas in ideal solids, stresses persist indefinitely. The higher this relaxation time, the more 'solid-like' the liquid behaves.

\begin{figure}\centerline{\includegraphics[width=87mm, clip,
keepaspectratio]{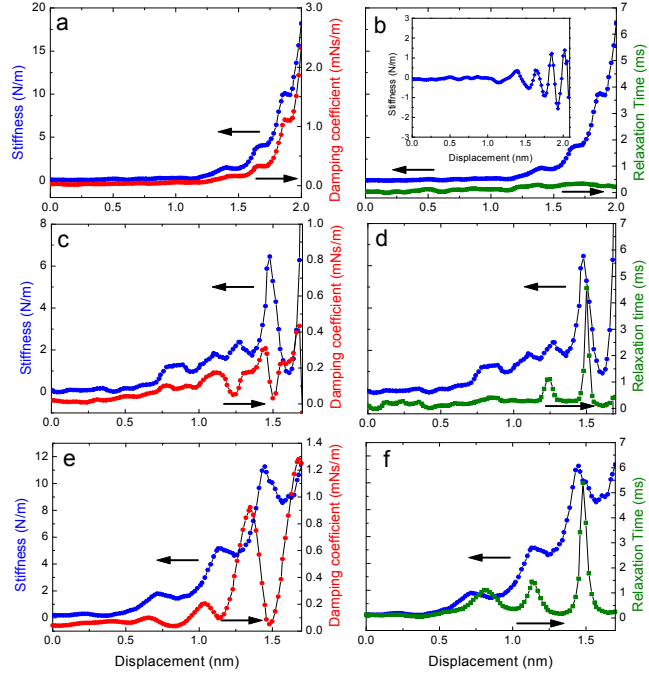}} \caption{(color online) Left column: Stiffness (blue) and damping coefficients (red) measured for the last 3-4 layers adjacent to the mica substrate versus sample displacement. Substrate is located to the right in each case. Right column: Stiffness and mechanical relaxation time (green) versus sample displacement. a and b: Approach speed 2 $\textrm{\AA}$/s, cantilever stiffness 2.4N/m; c and d: 8 $\textrm{\AA}$/s, 1.4 N/m. e and f: 14 $\textrm{\AA}$/s, 1.4 N/m. Inset to Fig. 1b: Stiffness after subtraction of repulsive background.}\label{stiffdamp}
\end{figure}

Figures \ref{stiffdamp}a-f shows measured stiffness, damping coefficient and mechanical relaxation time versus sample displacement. The mica surface is located to the right (at high sample displacements). In all cases, the observed oscillations in the stiffness and damping correspond to ordering of molecular layers of water. At certain tip-surface separations, an integer number of molecular layers can be accommodated in the gap. In these cases, we expect to measure higher stiffness. When the tip-surface gap is not commensurate with the molecular size, ordering of the molecules is frustrated and we observe reduced stiffness. Thus the stiffness oscillates with a period corresponding to the molecular spacing of the water layers. The average spacing of the layers we observed was 2.9 $\pm$ .8 $\textrm{\AA}$. Figures \ref{stiffdamp}a and b correspond to an approach speed of 2 $\textrm{\AA}$/s. At this speed, we observed weak oscillations in the stiffness (Fig. \ref{stiffdamp}a) superimposed on a repulsive, hydrophilic background. The inset to Fig. \ref{stiffdamp}b shows the stiffness after subtraction of the hydrophilic, repulsive background, showing clearly the range of ordering.The damping coefficient also shows oscillations, which are \emph{in-phase} with the oscillations in stiffness. We found the magnitude of the stiffness, $k$, depends on the tip radius, $R$. The stiffness can be normalized using the Derjaguin approximation \cite{der34} and corresponds to a maximum stress of about $k/R \approx 200 \textrm{ MPa}$. We imaged the tips before and after each measurement run using an electron microscope. However, due to observed tip changes during the measurements, we cannot determine the tip radius with certainty. This is not crucial to our discussion of the dynamical behavior of the water layers, however, because the mechanical relaxation time is independent of tip size, as it is calculated from the ratio of stiffness and damping coefficient.

Figure \ref{stiffdamp}b shows the relaxation time (and the stiffness for comparison) as a function of sample displacement. The relaxation time shown in Fig. \ref{stiffdamp}b exhibits no systematic change with increasing confinement or with the observed ordering. Thus, while both stiffness and damping oscillate weakly and monotonically increase, the liquid retains its viscous, liquid-like mechanical behavior under static to mildly dynamic conditions. The same results are obtained at approach speeds 4 $\textrm{\AA}$/s and 6 $\textrm{\AA}$/s (not shown). However, as shown in Fig. \ref{stiffdamp}c, the situation changes dramatically at an approach speed of 8 $\textrm{\AA}$/s (0.31 s per molecular layer). At this approach speed,  the damping coefficient switches from in-phase to out-of-phase with the stiffness as the tip-surface gap is reduced to two layers or less. The fact that the damping is \emph{reduced} when the stiffness is increased, means that when the liquid is ordered (i.e. when the gap is an integer multiple of the molecular size), the liquid has a strong elastic response (stiffness), but a weak dissipative response (damping). This suggests that the last two layers act almost purely elastically, i.e. mechanically, the liquid behaves "solid-like" in this regime. As far as we know, this is a new observation in nanoconfined water layers. The fact that the stiffness and damping are out-of-phase close to the mica surface leads to sharp peaks in the mechanical relaxation time (Fig. \ref{stiffdamp}d). This results in the final layer exhibiting an order-of-magnitude increase in relaxation time above the bulk value. As we increase the approach speed further, the out-of-phase (and, therefore, solid-like) behavior extends further into the bulk liquid. At 14 $\textrm{\AA}$/s (Fig. \ref{stiffdamp}e), the stiffness and damping are out-of-phase even for four confined layers, leading to prominent peaks in the relaxation time (Fig. \ref{stiffdamp}f).

\begin{figure}\centerline{\includegraphics[width=74mm, clip,
keepaspectratio]{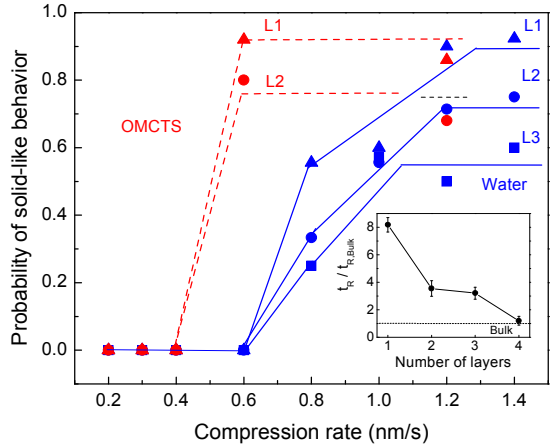}} \caption{(color online) Probability that layers show solid-like mechanical behavior. Blue: Water with 1 (L1) to 3 (L3) confined layers (L3). Red: OMCTS with 1 (L1) to 2 (L2) confined layers. Lines guide the eye and do not imply any functional relationship. Inset: Mechanical relaxation time versus number of confined layers in water, averaged for all rates above 8 $\textrm{\AA}$/s. Relaxation time strongly depends on layer number as shown, but is independent of squeeze rate as long as dynamical solidification is observed.}\label{prob}
\end{figure}

As an overview, we calculated the probability (from a total of 83 measurements) that the last three layers before contact exhibit elastic behavior. Figure \ref{prob} shows a sharp transition between bulk liquid-like behavior below 8 $\textrm{\AA}$/s to about 50\% solid-like behavior at 8 $\textrm{\AA}$/s when the water layer is squeezed down to a single molecular layer. As the speed is increased, the probability that a single layer behaves solid-like increases to almost 100\%. The graph also shows that the probability is lower for films of two or three molecular layers thickness, i.e. the probability of solid-like behavior increases both with approach speed and thickness of the film. Dynamic solidification is only observed for the last few molecular layers, i.e. when the film is less than 1 nm in thickness.

For comparison, the dynamic solidification of octamethylcyclotetrasiloxane (OMCTS), a silicone oil, is also shown \cite{pat06,zhu03}. OMCTS solidifies at even lower approach speeds, and the transition is sharper than in water. The fact that both OMCTS and water show this behavior is not necessarily to be expected, as water is dominated by hydrogen bonding, has faster molecular relaxation and a much lower melting temperature. The lower speed at which the transition occurs in OMCTS may be related to the higher viscosity of OMCTS, and the increased sharpness of the transition may correspond to the larger size of the molecules (making the ordering less vulnerable to the effects of atomic-scale surface roughness).

The relaxation times are summarized in the inset to Fig.\ \ref{prob}. This graph shows the relative increase in relaxation time, i.e. it shows by what factor the relaxation time increases above the bulk value upon dynamic 'solidification'. We found that once solidification is observed, the relaxation time is independent of approach speed, but depends strongly on film thickness.  It can be seen that in for 2-3 molecular layers, the relaxation time increased by a factor of 3, while for a single layer, the increase is almost an order of magnitude. Comparing these results to OMCTS \cite{pat06}, we find that water has a similar mechanical relaxation time, but OMCTS exhibited a larger relative increase of the relaxation time upon ordering (up to $20-30\times$ the bulk value). This may reflect the better ordering of the larger and more 'sluggish' OMCTS molecules.

When a liquid is squeezed out between a spherical tip and a flat surface, we expect squeeze damping due to the finite viscosity of the liquid. Squeeze damping is given by $\gamma_s = 6 \pi \eta \frac{R^2}{h} f^{*}$, where $\eta$  is the viscosity, $R$ is the tip radius, $h$ is the distance from the surface, and $f^{*}$ is a dimensionless factor of order 1 which describes the slip along the boundary ($f^{*}=1$ in the case of no slip) \cite{bon03}. Substituting reasonable values for the last molecular layer ($\eta= 10^{-3}$ Pa s, $R$ = 100 nm, $h$ = 0.25 nm, and $f^{*}=1$ ), we expect  $\gamma_s = 7.5 \times 10^{-7}$ Ns/m. This is several orders of magnitude lower than observed in our measurements. Other experiments also reported large increases in the effective viscosity \cite{ant01,li07}. However, we found a difference between measurements at slow and fast approach speeds, which have not been reported before. At slow speeds, the damping increases to $10^{-2}$ Ns/m, but at higher speeds, where solidification is observed, the damping is reduced to as low as a few $10^{-5}$ Ns/m when the liquid orders. Thus the increase in effective viscosity is largest at slow approach speeds. At faster compression rates, the liquid responds elastically in the ordered state, exhibiting high elastic stiffness and low damping.

Raviv et al.\ \cite{rav01} reported no significant changes in the viscosity of nanoconfined water, in contrast to other reports \cite{ant01, jef04, li07}. In their measurements, water was squeezed to a thickness of 1 nm. Then a sudden jump occurred once the pressure on the remaining water layer exceeded its yield strength. Our measurements agree that at 1 nm distance from the surface, confinement effects seem to be weak. Significant changes in behavior are only observed in the last 3-4 water layers \cite{ver06}. We also found that at high strain rates the effective damping coefficient and therefore the effective viscosity is reduced, although not as low as the bulk value. In the snap-in measurements of Raviv et al., high strain rates may have reduced the effective viscosity and these measurements, therefore, could not probe the quasistatic mechanical properties of the last 3-4 molecular layers.

We also compared our findings to shear measurements in water layers \cite{li08}. Although we were not shearing the layers, a squeeze-out of the liquid leads to lateral flow and therefore shear. Geometric arguments allow us to estimate the resulting shear strain rate as follows: The volumetric flux out of the gap under the tip is given by $j=\frac{1}{A}\frac{dV}{dt}=\frac{1}{2\pi R h}\pi R^2\frac{dh}{dt}=\frac{R}{2h}\frac{dh}{dt}$
This flux is equal to the flow speed of squeezed-out water. This leads to the strain rate: $\dot{s}=\frac{j}{h}=\frac{R}{2h^2}\frac{dh}{dt}$.

For a single molecular layer, the strain rate is about 160 for an approach speed of 2 $\textrm{\AA}$/s and 1120 for 14 $\textrm{\AA}$/s (assuming $R$ = 100 nm, $h$ =0.25 nm) . These rates are comparable to the strain rates measured by Li et al.\ \cite{li08}. They found a sharp increase in the shear relaxation time once they reduced the strain rate below about 500, which would correspond to an approach speed of about 6 $\textrm{\AA}$/s in our measurements. This is in reasonable agreement with our findings that dynamic solidification occurs above 6 $\textrm{\AA}$/s in water.

Li et al.\ compared the behavior of nanoconfined water to the nonlinear viscoelastic behavior of complex fluids. Clearly, the continuum picture of the liquid breaks down at this scale, and the liquid should be considered as a collection of interacting molecules, similar to large particles in a complex fluid. However, our observations do not fit neatly into this scheme. In shear-thickening liquids, for example, we would expect an increase in viscosity at high strain rates. In contrast, we observed a decrease in viscosity, accompanied by a large increase in elastic stiffness. In this sense, we observe a new phenomenon in water (and previously in OMCTS), which we call "dynamic solidification".

At low strain rates the molecules are able to easily diffuse out of the gap as the gap size is reduced (Fig. \ref{scheme}a). Both the stiffness and damping oscillate in unison with the number of molecular layers in the gap, indicating an oscillation of the confined liquid's density as the gap is decreased. However, at higher strain rates, the molecules cannot easily move out of the gap. This is likely due to the fact that they are in a very restricted volume and many molecules have to move collectively (Fig. \ref{scheme}b) \cite{pat06}. Thus molecules become "stuck" until the increasing pressure forces them out. During the small oscillation of the tip, the "stuck" molecular water layers respond elastically, i.e. the layers can store mechanical energy and release it back to the cantilever. However, the elastic response is only observed when the liquid is "ordered", i.e. when the gap is an integer multiple of the molecule size. In the "disordered" state (Fig \ref{scheme}a) at high strain rates, the liquid behaves liquid-like, but with enhanced viscosity, i.e. similar to shear-thickening in complex fluids.

In conclusion, the we show that, in water, both the elastic and viscous response oscillate with the molecular layering of the liquid as the gap is reduced, indicating that the both elasticity and viscosity of the liquid depend on the ability of the liquid to assume short-range order in the tip-surface gap. We also find that the mechanical response changes dramatically when the approach speed (strain rate) is increased. Above a certain threshold rate, the liquid behaves solid-like with low viscosity and high elasticity when the gap is commensurate with molecular size, while retaining a liquid-like, high viscosity state when the gap is incommensurate with the molecular size. These findings may explain some of the contradictory findings by other methods and may have important implications for nanofluidic systems and dynamics of macromolecular motion in cells.

\begin{acknowledgments}
P.\ M.\ H.\ would like to acknowledge funding through the National Science Foundation, grant DMR-0804283.
\end{acknowledgments}

\bibliography{waterspeed}

\end{document}